\documentclass[conference]{IEEEtran}
\usepackage{cite}
\usepackage{amsmath,amssymb,amsfonts}
\usepackage{algorithmic}
\usepackage{algorithm}
\usepackage{array}
\usepackage{graphicx}
\usepackage{textcomp}
\usepackage[caption=false,font=normalsize,labelfont=sf,textfont=sf]{subfig}
\usepackage{xcolor}
\usepackage[colorlinks,linkcolor=blue]{hyperref}
\begin{document}

\title{Design of a Decentralized Fixed-Income Lending Automated Market Maker Protocol Supporting Arbitrary Maturities}

\author{\IEEEauthorblockN{Tianyi Ma}
\IEEEauthorblockA{\textit{Antai College of Economics and Management} \\
\textit{Shanghai Jiao Tong University}\\
Shanghai, China \\
harryma-shs@sjtu.edu.cn
}
}

\maketitle

\begin{abstract}
In decentralized finance (DeFi), designing fixed-income lending automated market makers (AMMs) is extremely challenging due to time-related complexities. Moreover, existing protocols only support single-maturity lending. Building upon the BondMM protocol, this paper argues that its mathematical invariants are sufficiently elegant to be generalized to arbitrary maturities. This paper thus propose an improved design, BondMM-A, which supports lending activities of any maturity. By integrating fixed-income instruments of varying maturities into a single smart contract, BondMM-A offers users and liquidity providers (LPs) greater operational freedom and capital efficiency. Experimental results show that BondMM-A performs excellently in terms of interest rate stability and financial robustness.

\end{abstract}


\section{Introduction}
Following the 2008 financial crisis, the emergence and development of Bitcoin \cite{bitcoin} and blockchain technology have introduced disruptive changes to the financial sector. Bitcoin pioneered decentralized digital money, and Ethereum \cite{ethereum}, with its smart contract capabilities, further expanded blockchain applications in finance, catalyzing the rise of Decentralized Finance (DeFi) \cite{b1} and attracting massive participation.

In traditional finance, fixed-income instruments—commonly known as bonds—are financial assets that promise predetermined cash flows at specified future dates. They hold a dominant position in global financial markets: as of year-end 2023, global equity market capitalization stood at approximately \$115 trillion, while the global bond market reached \$141 trillion. Moreover, daily U.S. bond trading volume is roughly twice that of equities\footnote{Data Source: Huatai Xingzhi (\href{https://stock.jrj.com.cn/2024/12/03080145936337.shtml}{reprinted})}. Despite the DeFi lending market having grown to \$54.6 billion\footnote{Data Source: \href{https://defillama.com/protocols/Lending}{DefiLlama}}, truly fixed-income products remain rare \cite{aave}-\cite{notional}.

Building on the BondMM protocol \cite{bmm}, this paper proposes BondMM-A, a novel AMM-based smart contract design supporting arbitrary maturities for decentralized fixed-income lending, aiming to meet this unmet market demand.

\textbf{Related Work and Limitations:} In DeFi, lacking traditional financial intermediaries, lending is typically facilitated through Automated Market Makers (AMMs). An AMM is a smart contract that enables users to lend, withdraw loans, borrow, and repay. Liquidity Providers (LPs) initially fund the AMM’s liquidity pool, and all user interactions occur directly with this pool. LPs earn fees or interest spreads as compensation.

Dominant DeFi lending protocols—such as AAVE \cite{aave}, Compound \cite{compound}, and MakerDAO \cite{makerdao}—primarily employ variable interest rates, treating fixed rates as supplementary features. Dedicated fixed-income protocols include Yield Protocol \cite{yield} and Notional Protocol \cite{notional}.

Theoretically, these protocols set identical borrowing and lending rates, meaning LPs profit only from transaction fees. This requires sufficient secondary market trading (discussed below) rather than primary issuance. Speculators prefer longer-dated bonds due to higher leverage potential, yet current fixed-income AMMs offer only short maturities (maximum 1 year), limiting market depth. The Yield Protocol suffers from low capital efficiency and insufficient liquidity, often halting under stress. Notional Protocol attempts to resolve this but introduces new issues, such as excessively high pricing.

The BondMM protocol \cite{bmm} uses a set of simple, closed-form mathematical invariants to achieve fair pricing, high capital efficiency, and stable LP equity, offering a promising path forward. However, like others, it supports only single-maturity products per contract. Since each AMM requires separate LP funding, multi-maturity support necessitates multiple disjoint liquidity pools, exposing LPs to fragmented capital risk.

\textbf{Contributions:} This paper enhances BondMM to propose BondMM-A, which supports arbitrary maturities within a single contract. Users can now choose from a continuous range of maturities in one interface, increasing flexibility. LPs provide liquidity to a unified pool, eliminating capital fragmentation. Overall, BondMM-A integrates user and LP needs, improving operational efficiency and capital utilization, and offering an innovative solution for decentralized fixed-income markets.

The remainder of this paper is structured as follows: Section II introduces foundational concepts; Section III details the BondMM-A design; Section IV presents experimental evaluation; Section V concludes with future outlook.

\section{Preliminaries}

\subsection{Fixed-Income Instruments and Interest Rates}
Mathematically, a fixed-income instrument can be represented as a pair of sequences: $\{t_i\}$ and $\{C_{t_i}\}$, where $t_i$ means cash flow times and $C_{t_i}$ means cash flow amounts. Typically, $C_{t_1}<0$, and $C_{t_i}>0$ for $i>1$. This means investors pay an initial outflow to purchase the bond, and receive subsequent inflows without further payment. 

In the simplest case, $\{t_i\}=\{0,T\}$, $\{C_{t_i}\}=\{-P,F\}$, where $P$ is the bond price and $F$ is the face value. Example: A 2-year bond with $F = 100$, $P = 95$ yields $5$ in interest, giving a simple return of $5/95 \approx 5.26\%$. To distinguish different maturities, practitioners use the annualized rate $R=\frac{F}{P}^\frac{1}{2}-1\approx2.60\%$. This interprets the 2-year bond as two consecutive 1-year bonds: investing $P$ at the start of year one returns $P(1+R)$ at year-end; reinvesting at the same rate returns $P(1+R)^2$ at year two. If the annualized rate is constant over time, it satisfies $\sum\limits_{i}C_{t_i}(1+R)^{-t_i}=0$. In continuous compounding, let $r=\ln(1+R)$; then $\sum\limits_{i}C_{t_i}e^{-t_i r}=0$. In the example, $r=\ln(1+R)\approx2.56\%$. Unless noted, all rates below are continuously compounded annualized rates. Complex fixed-income products can be decomposed into simple ones via their cash-flow series, so this paper focus on the simplest case: $\{t_i\}=\{0,T\}$, $\{C_{t_i}\}=\{-P,F\}$.

In reality, market mechanisms equalize rates for products with the same maturity—the market rate—but it fluctuates over time, so different maturities carry different rates. In the earlier example, if both 1-year and 2-year rates are $2.56\%$ initially but the 1-year rate rises after one year, the optimal choice is two consecutive 1-year bonds. Strictly, the instantaneous rate $r$ is a continuous function $r(t)$; a simple product satisfies $F=Pe^{\int_0^Tr\text{d}t}$. Because $r(t)$ is uncertain, it is usually assumed constant over short horizons in practice.

Rate volatility makes bonds speculative assets. Traders can profit by buying and selling between issuance and maturity. Trades at issuance and maturity are primary trades that mint or burn bonds; trades in between are secondary trades that merely transfer bonds.

\subsection{Fixed-Income AMM}

Consider a bond maturing at time $T$ with face value $1$ (in some digital currency such as DAI); it can be tokenized. A fixed-income AMM holds two tokens: bonds and currency. Currency is supplied by LPs; bonds are minted when the AMM is created. At any time, let $t$ be time to maturity; the AMM state is $(x,y,r,p)$, where $x,y$ are bond and currency balances, $r$ is the marginal rate, and $p$ is the marginal price. Assuming no spread, $r$ is the marginal rate for both lenders and borrowers. Theoretically $p=e^{-tr}\Leftrightarrow r=\frac{1}{t}\ln\frac{1}{p}$. In practice AMMs define $r,p$ as functions of $(x,y)$ so market forces adjust them; $(x,y)$ uniquely determines state, and different designs give different formulas for $r,p$.

An AMM typically exposes four operations: lend (buy bonds), withdraw a loan (sell bonds), borrow, and repay. Each changes the two balances inversely. Lending injects cash and receives bonds; withdrawing injects bonds and receives cash. Borrowing requires collateral (often worth $150\%$ of the loan), receives cash, and the pool mints bonds; repaying injects cash, returns collateral, and the pool burns bonds. If a borrower fails to repay before maturity, the pool liquidates collateral. Any trade can be expressed as $(\Delta x,\Delta y)$: $\Delta x>0$ (i.e., $\Delta y<0$) for borrowing or withdrawing, and $\Delta x<0$ (i.e., $\Delta y>0$) for lending or repaying.

An ideal fixed-income AMM should satisfy:

\textbf{Property 1 (Operability):} After a trade $(\Delta x,\Delta y)$ on AMM $(x,y)$, balances remain nonnegative: $x+\Delta x\ge0,\;y+\Delta y\ge0$.

\textbf{Property 2 (Computability):} Given $\Delta x$, compute the unique $\Delta y$ (or prove it absent) in constant time, and vice versa.

\textbf{Property 3 (Par Redemption):} At $t=0$, $r=0,p=1$, independent of $x,y$; one unit of bond equals one unit of currency.

\textbf{Property 4 (Path Independence):} For fixed $t$, if $(\Delta x_1,\Delta y_1)$ and $(\Delta x_2,\Delta y_2)$ are two consecutive valid trades from state $(x,y)$, then $(\Delta x_1+\Delta x_2,\Delta y_1+\Delta y_2)$ is also valid. Equivalently, buying $\Delta x_1$ then $\Delta x_2$ units of bonds at the same moment equals buying $\Delta x_1+\Delta x_2$ once.

Path independence typically comes from a conservation relation $f_t(x, y)=0$ where, for fixed $t$, $x$ and $y$ map one-to-one in the domain. Any valid trade satisfies $f_t(x+\Delta x, y+\Delta y)=0$. Then $r=-\frac{\partial y}{\partial x}$ gives the marginal rate and $p=e^{-tr}$ the marginal price.

\textbf{Property 5 (Economic Soundness):} Convergence to market equilibrium should be stable with limited arbitrage. In particular, rates should be nonnegative, i.e., prices should not exceed 1.

\subsection{Existing Protocols}

This paper summarizes existing fixed-income AMMs: Yield, Notional, and BondMM.

\subsubsection{Yield Protocol}

The Yield Protocol \cite{yield} constructs the invariant:

$$f_t(x,y)=x^{1-t/T}+y^{1-t/T}-C(t)=0,$$

Therefore

$$p=-\dfrac{\partial y}{\partial x}=\left(\dfrac{y}{x}\right)^{t/T}=\phi^{-t/T},$$

And thus

$$r=\dfrac{1}{t}\ln\dfrac{1}{p}=\dfrac{1}{t}\ln\phi^{t/T}=\dfrac{1}{T}\ln\phi,$$

Here $\phi=\frac{y}{x}$ is the bond-to-cash ratio. When $\phi<1$, $r<0$ and negative rates appear, violating economic soundness; no one wants to lend, severely reducing liquidity.

\subsubsection{Notional Protocol}

Unlike the general case, the Notional Protocol \cite{notional} prices with simple interest, $p(1+rt)=1$, where $t$ is time to maturity. To avoid Yield's negative-rate issue, Notional introduces two (possibly time-varying) parameters $\kappa\in(0,1)$ and $r^*\in(0,1)$, where $r^*$ is the anchor rate and $\kappa$ controls rate volatility. Under this protocol,

$$r=\kappa\ln\phi+r^*,$$

So

$$p=\dfrac{1}{1+rt}=(1+t\kappa\ln\phi+tr^*)^{-1}.$$

Notional does not price via a conservation law but via a pricing formula. Given $\Delta x$, define

$$\bar\phi=\dfrac{x+\Delta x}{y-\Delta x},$$

The average trade price is

$$\bar p=(1+t\kappa\ln\bar\phi+tr^*)^{-1},$$

So $\Delta y=-\bar p\Delta x$.

If $\Delta x<0$ (lend or repay), then $\bar p<1$, so the post-trade $\phi^\prime=\frac{x+\Delta x}{y-\bar p\Delta x}<\bar\phi$, giving $p^\prime>\bar p$. The user pays a price higher than both surrounding marginal prices—an extra interest cost to LPs—so Notional is less attractive to borrowers. It clearly lacks path independence. It is also hard to derive $\Delta x$ from $\Delta y$, violating computability.

\subsubsection{BondMM Protocol}

The BondMM protocol \cite{bmm} introduces the bond present value (current-time price)

\begin{equation}\label{X}
    X=xp=xe^{-rt},
\end{equation}

And sets the rate as a function of $X/y$:

$$r=R(\psi),\psi=\dfrac{X}{y},$$

Unlike protocols that keep the bond amount fixed over time, BondMM seeks—by economic intuition—to keep the present value of bonds unchanged. With positive rates, present value decays, so BondMM gradually burns bonds in the pool.

Specifically, BondMM chooses $R$ similar to Notional:

\begin{equation}\label{r}
    r=R(\psi)=\kappa\ln\psi+r^*=\kappa\ln\dfrac{X}{y}+r^*.
\end{equation}

It can be shown (Tran et al. \cite{bmm}, likewise below):

$$r=\dfrac{1}{1+\kappa t}\left(\kappa\ln\dfrac{x}{y}+r^*\right),$$

$$p=\left[\left(\dfrac{x}{y}\right)e^{r^*}\right]^{-t/(1+\kappa t)}.$$

Assume at pool creation LPs deposit $y_0$ cash and the initial rate is $r_0$. To satisfy $r_0=r^*$,  $X_0=y_0$ is needed, so mint $x_0=X_0/p_0=X_0e^{Tr_0}$ units of bonds.

At any moment,

$$p=-\dfrac{\partial y}{\partial x}=e^{-rt},$$

And thus

$$r=R\left(\dfrac{X}{y}\right)\Rightarrow y=\dfrac{X}R^{-1}(r)=\dfrac{xe^{-rt}}{R^{-1}(r)},$$

With initial values $(x_0,y_0,r_0)$, solving the differential equations yields:

$$\begin{array}{l}x=X_0e^{r_0}\left[e^{(r-r_0/\kappa)}\dfrac{e^{(r_0-r^*)/\kappa}+1}{e^{(r-r^*)/\kappa}+1}\right]^{1+\kappa t},\\y=y_0\left[\dfrac{e^{(r_0-r^*)/\kappa}+1}{e^{(r-r^*)/\kappa}+1}\right]^{1+\kappa t}\end{array},$$

Hence BondMM satisfies the invariant

\begin{equation}\label{xy}
    Kx^\alpha+y^\alpha=C,
\end{equation}

Where

\begin{equation}\label{akc}\alpha=(1+\kappa t)^{-1},K=e^{-tr^*\alpha},C=y_0^\alpha[e^{(r_0-r^*)/\kappa}+1],\end{equation}

And $X$ and $y$ also satisfy an invariant:

\begin{equation}\label{Xy}y^\alpha(\dfrac{X}{y} + 1)=C.\end{equation}

For a new trade $(\Delta x,\Delta y)$, the AMM state before and after satisfies the invariant \eqref{xy}:

$$K(x+\Delta x)^\alpha+(y+\Delta y)^\alpha=C=Kx^\alpha+y^\alpha,$$

Thus, the following can be calculated from $\Delta x$:

\begin{equation}\label{dy}
    \begin{split}
        \Delta y&=\left[C-K(x+\Delta x)^\alpha\right]^{1/\alpha}-y\\
        &=y\left[\dfrac{X}{y}+1-\left(\left(\dfrac{X}{y}\right)^{1/\alpha}+e^{-r^*t}\dfrac{\Delta x}{y}\right)^\alpha\right]^{1/\alpha}-y,
    \end{split}
\end{equation}

Similarly, the following can be calculated from $\Delta y$:

\begin{equation}\label{dx}
    \Delta x=e^{r^*t}y\left[\left(\dfrac{X}{y}+1-\left(\dfrac{\Delta y}{y}+1\right)^a\right)^{1/\alpha}-\left(\dfrac{X}{y}\right)^{1/\alpha}\right].
\end{equation}

In Tran et al.'s experiments \cite{bmm}, $r^*$ is set to $r_0$ and $\kappa$ to $0.02$, but both can vary over time so BondMM can adapt to market conditions. With suitable parameters, BondMM satisfies operability, computability, par redemption, path independence, and economic soundness.

BondMM also considers financial stability: the pool's cash should cover all debt, so trades that could cause insolvency are rejected. Let $b$ and $l$ denote borrow and lend amounts; net equity is $E=y+(b-l)p$. Ideally $E=y_0$, meaning LPs neither lose nor gain beyond fees. BondMM rejects lending when $E$ trends substantially below $y_0$ \cite{bmm}.

\section{BondMM-A: New Design}

BondMM replaces face value with present value in the time conservation, inspiring a new design path. This will bring a question: Is tokenizing present value instead of the bond itself be a better way to further simplify fixed-income AMMs and support arbitrary maturities? Fortunately, BondMM's elegant invariants enable this.

From \eqref{X} and \eqref{r}:

\begin{equation}\label{x}
    x=Xe^{rt}=Xe^{r^*t+\kappa t\ln(X/y)},
\end{equation}

Because the BondMM-A pool itself has no maturity, AMM state is no longer tied to clock time. Here $t$ denotes the maturity of the bond under consideration, and $x$ is the face value of a $t$-maturity bond equivalent to the pool's remaining present value. The parameters $\alpha,K,C$ in \eqref{akc} are functions of $t$, so the invariants \eqref{xy} and \eqref{Xy} still hold with the new meaning of $t$, yielding infinitely many invariants at each moment. After a trade with maturity $t$, only one invariant remains valid; others break. This does not affect operability, computability, par redemption, path independence, or economic soundness. Pricing still follows \eqref{dy} and \eqref{dx}, with $t$ being the trade's maturity. Pool initialization matches BondMM, with required face value $X_0=y_0$ minted.

For implementation, trades specify both amount and maturity, so transferring “present-value tokens” alone is insufficient. Alternatives include recording each user's bond face value and maturity on-chain, or restricting maturities to a finite set and issuing one token per discrete tenor.

If $r^*$ is fixed, BondMM-A would produce identical rates across maturities, contradicting reality. Since $r^*$ can depend on maturity, the contract can set $r^*$ as a function of tenor (e.g., a polynomial) and let the owner tune parameters to shape the curve and match market shifts.

Like BondMM, BondMM-A must ensure solvency. Because trades span different maturities, simply computing $E=y+(b-l)p$ is inefficient. Instead maintain $L$, the present value of all outstanding borrows minus unrepaid loans, updating it over time via $\text{d}\ln L=r\text{d}t$. Then $E=y+L$; if $E$ trends meaningfully below $y_0$, BondMM-A rejects lending.

\section{Experimental Evaluation}

This paper evaluates BondMM-A\footnote{Code: \href{https://github.com/HarryTMa/BondMMA}{https://github.com/HarryTMa/BondMMA}} focusing on effectiveness (do rates match market rates?) and stability (is net equity stable?).

\textbf{Parameter setup:} The pool starts with cash $y_0=1000$, initial rate $r_0=5\%$, $\kappa=0.02$, and a common $r^*$ across maturities. The market curve is generated by the Cox-Ingersoll-Ross (CIR) model \cite{cir} with $k=0.4,\theta=0.05,\sigma=0.2$. The horizon is $T=1$ year, divided into $N=100000$ steps; each step launches $M=1000$ active trades. At each step end, set $r^*$ to the previous step's market rate. Active trades follow speculation: if BondMM-A's rate exceeds the market rate, users lend; otherwise they borrow. Each trade's maturity (absolute value of $N(T-t,T-t)$ with current time $t$) and size (absolute value of normal distribution $N(0.72,1)$, expected near $1$) are random. Loans and borrows maturing within $T$ are settled as passive trades. If active and passive trades coexist, they interleave evenly (e.g., $1000$ active and $500$ passive imply 2 active then 1 passive). Lending stops if $E$ falls below $99\%$ of $y_0$ at any time.

\begin{figure}
    \centering
    \includegraphics[width=0.5\textwidth]{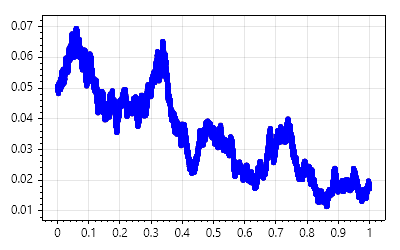}
    \caption{Comparison of BondMM-A average rate and market rate; they nearly overlap.}
    \label{fig:r}
\end{figure}

\begin{figure}
    \centering
    \includegraphics[width=0.5\textwidth]{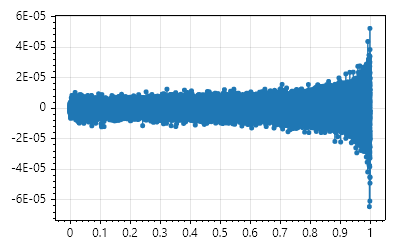}
    \caption{Difference between BondMM-A average rate and the market rate.}
    \label{fig:diff}
\end{figure}

\begin{figure}
    \centering
    \includegraphics[width=0.5\textwidth]{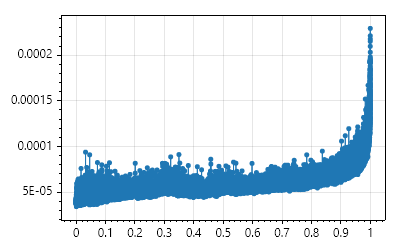}
    \caption{Standard deviation of BondMM-A rates at each time step.}
    \label{fig:std}
\end{figure}

\begin{figure}
    \centering
    \includegraphics[width=0.5\textwidth]{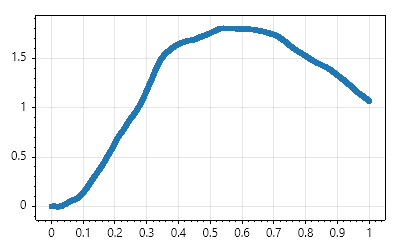}
    \caption{BondMM-A net equity minus initial capital.}
    \label{fig:e}
\end{figure}

As Figure \ref{fig:r} shows, BondMM-A rates closely track market rates, demonstrating effectiveness. Figures \ref{fig:diff} and \ref{fig:std} show the rate gap is around $10^{-5}$ and the standard deviation around $10^{-4}$—both tiny; although they rise slightly later, magnitudes stay small. BondMM-A is effective.

Figure \ref{fig:e} shows net equity: BondMM-A stays about $1$ above the initial $1000$, indicating strong financial stability. Because equity never dropped below $99\%$ of initial capital, no loan refusals occurred.

\section{Conclusion and Outlook}

This paper proposed BondMM-A, a fixed-income AMM supporting arbitrary maturities. Its design proves effective: rates align with market levels, volatility is low, net equity is stable, users gain flexibility, and LPs avoid capital fragmentation—benefiting both sides.

BondMM-A still invites optimization. Future work includes more precise dynamic parameter tuning for fast markets, richer yield-curve models to match real-world diversity, and stronger resilience to extremes. Real deployments will test performance and security at DeFi scale. Overall, BondMM-A offers a new design path for decentralized fixed income and a reference for subsequent research.


\begin{thebibliography}{00}
\bibitem{bitcoin} Nakamoto S. Bitcoin: A peer-to-peer electronic cash system[J]. Satoshi Nakamoto, 2008.
\bibitem{ethereum} Buterin V. A next-generation smart contract and decentralized application platform[J]. Ethereum White Paper, 2014, 1(1): 1-32.
\bibitem{b1} Schueffel P. Defi: Decentralized finance-an introduction and overview[J]. Journal of Innovation Management, 2021, 9(3): I-XI.
\bibitem{aave} Frangella E, Herskind L. Aave v3 technical paper[J]. 2022.
\bibitem{compound} Leshner R, Hayes G. Compound: The Money Market Protocol—Whitepaper[J]. 2019.
\bibitem{makerdao} Team M. The Maker Protocol: MakerDAO's Mulfi-Collateral Dai (MCD) System[J]. 2019.
\bibitem{yield} Niemerg A, Robinson D, Livnev L. YieldSpace: An automated liquidity provider for fixed yield tokens[J]. 2020.
\bibitem{notional} Notional Finance. Notional Whitepaper[J]. 2020.
\bibitem{bmm} Tran T, Tran D A, Truong K. Financially-Stable Automated Market Making for Decentralized Fixed-Rate Lending and Trading[C]//2024 IEEE International Conference on Blockchain and Cryptocurrency (ICBC). IEEE, 2024: 353-361.
\bibitem{cir} Cox J C, Ingersoll Jr J E, Ross S A. A theory of the term structure of interest rates[J]. Econometrica, 1985, 53(2): 385-407.

\end{thebibliography}
\end{document}